# Fusion and Fission of Particle-like Chiral Nematic Vortex Knots

Darian Hall[1], Jung-Shen Benny Tai,[1] Louis H. Kauffman[2,3] and Ivan I. Smalyukh[1,3,4,5*]

[1]Department of Physics, University of Colorado, Boulder, CO 80309, USA
[2]Mathematics Department, 851 South Morgan Street, University of Illinois at Chicago, Chicago, Illinois 60607-7045, USA
[3]International Institute for Sustainability with Knotted Chiral Meta Matter, Hiroshima University, 1-3-1 Kagamiyama, Higashi-Hiroshima, Hiroshima 739-8526, Japan
[4]Materials Science and Engineering Program, University of Colorado, Boulder, CO 80309, USA
[5]Renewable and Sustainable Energy Institute, National Renewable Energy Laboratory and University of Colorado, Boulder, CO 80309, USA
* Correspondence to: ivan.smalyukh@colorado.edu

**Vortex knots have been seen decaying in many physical systems. Here we describe topologically protected vortex knots, which remain stable and undergo fusion and fission while conserving a topological invariant analogous to that of baryon number. While the host medium, a chiral nematic liquid crystal, exhibits intrinsic chirality, cores of the vortex lines are structurally achiral regions where twist cannot be defined. We refer to them as "dischiralation" vortex lines, in analogy to dislocations and disclinations in ordered media where, respectively, positional and orientational order is disrupted. Fusion and fission of these vortex knots, which we reversibly switch by electric pulses, vividly reveal the physical embodiments of knot theory's concepts like connected sums of knots[3]. Our findings provide insights into related phenomena in fields ranging from cosmology to particle physics and can enable applications in electro-optics and photonics, where such fusion and fission processes can be used for controlling light.**

Lord Kelvin's attempts to develop physics models of chemical elements led to the modern-day knot theory,[1-4] a branch of pure mathematics, as well as to concepts of chirality and topology that play essential roles across the entire nature's hierarchy, from elementary particles to soft, biological and quantum matter and to cosmology[5-17]. Fascinating experimental analogues of Kelvin's vortex knot models of atoms were recently studied in common media like water[12], but complex knots were found to decay to simpler counterparts and disappear after a series of reconnections of the vortex lines, so far finding no technological utility. On the other hand, liquid crystals (LCs) are known for their widespread applications, ranging from information displays to soft robotics and biodetection[18-28]. However, their technological utility mainly relies on continuous deformations of the orientational order of rodlike molecules in response to fields and other stimuli[20-27], even though topological defects are often used in some functionality designs, like mechanical actuation, guided nanoscale self-assembly and beam steering[19,23,28,29]. At the same time, recent developments in nematic colloids and chiral LCs allowed one to obtain controllably realized closed loops and knots of vortex lines and particle-like topological knot solitons stabilized by surface boundary conditions on colloidal surfaces or by medium's chirality in the bulk of chiral media[10,30-39]. However, the possibilities of using external stimuli for inducing fusion, fission and various reconnections of such topological objects, including inter-transformations between distinct states, as well as dynamics of such processes have not been studied, albeit control of particle-induced knots of disclination defects by laser tweezers was demonstrated.[32] Could electric switching of such fascinating topological objects further enhance the vast electrooptic technological potential of LCs, in addition to providing vivid demonstrations and experimental tests of the mathematical knot theory at work? Towards this goal, we explore how low-voltage electric fields can guide controlled transformations of stable Kelvin-atom-like vortex knots in chiral LCs through fusion, fission and more complex re-linking of knots.

Fission and fusion of atoms release massive amounts of energy while the net total number of nucleons, protons and neutrons, is conserved. Anyons in quantum computing[40], skyrmions in optics[41], and many other particles and topological quasi-particles exhibit similar processes, but they are hidden from direct experimental observations and often difficult to control. We describe how topologically protected vortex knots in the chiral LC medium undergo directly observable reconnections while conserving integer-valued topological invariants, mimicking nuclear fusion and fission. Much like in subatomic systems, our soft matter analogues of fusion and fission always lead to lower energy of the final state. Interestingly, pulses of electric field can controllably alter the energetics of these states and sequentially fuse or split the same particle-like vortex knots, which would be impossible to achieve with subatomic-physics counterparts of our knotted particle-like objects. The facile control of such localized knotted structures in the chiral LC's helical axis field promise knot-theory-guided photonic and electro-optic applications and unconventional computation, as well as data storage and spintronics applications for topologically similar knots realized in magnetic systems.[9-11]

**Electric-pulse controlled interactions between knots**

Our chiral nematics are confined in a geometry similar to that of electro-optic devices and displays[22,24] (Extended Data Fig. 1a-c), where the chiral LC is sandwiched between transparent indium tin oxide (ITO) electrodes, to which a 1 kHz alternating-current (AC) voltage is applied. Far away from the knots, the helical axis $\chi(\mathbf{r})$, around which molecules twist, is spatially uniform and orthogonal to confining substrates. Localized vortex knots in $\chi(\mathbf{r})$, regions where directionality of twist cannot be defined, are obtained by locally melting and quenching the LC using laser tweezers incorporated into an inverted microscope imaging setup. These localized knots are the so-called heliknotons, topological solitons with the hopfion topology in the material director field **n(r)**, but exhibiting singular vortex lines in $\chi$(**r**) (Extended Data Fig. 1)[10]. The stability of heliknotons in the LC mixture used in this study depends strongly on the magnitude of the applied voltage, expanding as the voltage increases and shrinking with reduced voltage, also stabilizing heliknotons of larger Hopf index at lower voltages and elementary ones at higher voltages. These dynamical properties of heliknotons allow us to finely tune vortex knot interactions through applying electrical pulses or continuously changing voltage.

While the connected sum of knots is a pure mathematical concept of reconnecting strands of two different knots through the so-called band surgery operation (Fig. 1a), similar reconnections also emerge in biological contexts[42], preserving the number of under/over crossings within the ensuing composite knot. Our vortex knots with disrupted twisting in their cores (Fig. 1b) commonly also exhibit more complex types of fusion, reconnecting simultaneously at two connecting sites of interacting knots (Fig. 1c-g, SI Video 1). The vortex lines forming knots can have locally defined winding numbers of 1/2 or -1/2 (Fig. 1c,g), characterizing local cross-sections of these defects, where here the winding numbers quantify the angle by which $\chi$(**r**) rotates around the vortex line when one navigates around its core once, divided by $2\pi$. During the reconnection at two sites simultaneously, the fragments of vortex lines of opposite winding number ($\pm 1/2$) annihilate and effectively lead to two band surgeries (Fig. 1c-g). Such transformations first lead to a two-component link and then, with the subsequent reconnections, to a three-component link (Fig. 1c). To gain insights into how these transformations take place in our chiral nematic system, we visualize the smoothly vectorized **n(r)** with the help of its color-coded order-parameter space, the 2-sphere $\mathbb{S}^2$ (Fig. 1d,e). This analysis reveals that **n(r)** stays continuous during such reconnections as the knots approach and fuse, having cores of merging vortex lines exhibit locally the same $\mathbf{n}(\mathbf{r})$ orientation. To reveal the behaviour of singular $\chi(\mathbf{r})$ during reconnection, we also visualize regions with significant bend and splay distortions in $\chi(\mathbf{r})$ (Fig. 1f). This reveals the fine details of local annihilation of opposite-winding-number vortex line fragments, resulting in the reconnection, where the $\pm 1/2$ winding number of a given vortex fragment is encoded in the bend-splay pattern (Fig. 1g).

Polarizing optical microscopy (POM), coupled with numerical POM and free-energy modelling of the system (Fig. 2a-g, SI Video 2), reveals how fusion of individual heliknotons progresses upon changing the applied voltage. The evolution of the separation vector (Fig. 2e) tracks the dynamics of the heliknotons as they fuse together. The relinking pathways extracted from experimental images and from energy-minimizing evolution of the $\mathbf{n}(\mathbf{r})$ and $\chi(\mathbf{r})$ fields closely match (Fig. 2f,g and SI Video 3). Interestingly, relinking, both within an individual heliknoton and between an interacting pair, can be driven along different kinetic pathways (Fig. 2h,i), which can be understood in terms of possible band surgeries between the vortex line segments within individual knots or their pairs. In addition to the various types of double-reconnections (Figs. 1 and 2), the reconnections representing more classical analogues of mathematical "connected sum of knots", schematically shown in Fig. 1a, are observed when heliknotons approach each other with the separation vector parallel to the far-field helical axis $\chi_0$ (Fig. 3a,b and SI Video 4), which can be induced by sub-second pulses of the electric field. The relinking response times $\tau_a$ and $\tau_o$, defined as times needed for the fusion- or fission-type knot relinking to occur after the electric field is turned on or off, respectively, also occur in the sub-second range (Fig. 3b). Relinking times for double reconnection and reconnections of vortex lines within individual heliknotons are also characterized by times in the sub-second to second range, albeit somewhat longer pulses are typically needed to prompt these topological transformations (Fig. 3c,d and SI Video 5). Response times can be tuned by controlling the electric field pulse amplitude whereby $\tau_a$ ($\tau_o$) can be shortened by increasing (reducing) the strength of the pulse (see Extended Data Fig. 7f,g). While the band surgeries associated with knot transformations are classified as incoherent[3,43-45] in nature in most cases, the examples shown in Fig. 3a,b can be considered as physical manifestations of the mathematically coherent (preserving orientation) band surgery[2,3], with oriented vortex constituent knots undergoing fusion.

**Complex knots, graphs, and analogues of high-baryon-number atoms from heliknoton fusion**

Beyond single- and multi-component knots and links, large composite structures formed via fusion of many separate knots also exhibit topological features of graphs, which in our case are structures composed of edges in the form of vortex lines and vertices at their junctions. While 3D spatial graphs are commonly seen as transient states separating the distinct knots before and after re-linking (Figs. 1-3), for large structures formed via fusion of many knots they also emerge as energy-minimizing or metastable states (Fig. 4c-j, SI Videos 6-10). Visualization of $\mathbf{n}(\mathbf{r})$ in the zoomed-in regions of complex inter-vortex's junctions forming graphs confirms that the material director field remains nonsingular within them (Fig. 4), as well as illustrates details of branching, connections, and re-connections of the vortex lines in the $\chi(\mathbf{r})$-field. Some of the components of complex vortex knots have multiple transformations between the local 1/2 and -1/2 structures along the closed loops, whereas the other components can maintain a single winding number, corresponding to 1/2 or -1/2 (Figs. 1c-g, Fig 2h,i).

While fusion and fission of two elementary heliknotons already has many diverse scenarios (Figs. 1-3), dependent on the relative directionality of fusion of knots, even more possibilities arise for such processes involving more than two heliknotons (Fig. 4 and Extended Data Figs 2-4). The pathways of fusion of knots can be controlled by tuning the applied voltage and using laser tweezers, where the latter allows us to spatially translate heliknotons and to locally melt or realign the chiral LC in-between the vortex knots, thus prompting the desired reconnections to occur. Within the interior of a single heliknoton, the knotting topology can be controlled by changing the applied voltage. In particular, the trefoil vortex state transforms to a three-component link by reducing the voltage (Fig 2h). The corresponding reverse transformation can be induced by increasing the voltage. Proximity to other heliknotons can influence the interior knotting even in the absence of inter-heliknoton relinking, in some cases, producing the Solomon link (Fig 2h) due to attractive heliknoton-heliknoton interactions. At the first sight, the relation between the complex diverse knots and the concept of quasi-atoms is elusive because the numbers of components (vortex loops and knots) as well as the numbers of crossings change during successive reconnections related to fusion and fission and other knot transformations. One would expect having integer invariants characterizing the quasiparticle knots that could represent the effective "baryon numbers" or the effective number of nucleons. Despite seemingly infinite large spectrum of possibilities of re-linking processes that do not exhibit conserved knot topology, we find that the cumulative Hopf index of heliknotons, defined in the material director field **n(r)**, is conserved during the fusion, fission and various re-linking transformations. Indeed, characterizing the Hopf index as an integral, we find that the Hopf indices follow the addition of the components during fusion and stay conserved as the number of elementary heliknotons taking part in interactions/transformations (Extended Data Figs. 5 and 6). For a solitonic unit vector field **n(r)** embedded in $\mathbb{R}^3$ whose far-field background allows for compactification into $\mathbb{S}^3$, the Hopf index $Q$ can be obtained as[38,47]

$$Q = \frac{1}{64\pi^2} \int_{\mathbb{R}^3} d^3\boldsymbol{r}\, \epsilon^{ijk}\, A_i F_{jk}\,, \quad (1)$$

where $F_{ij} = \epsilon_{abc} n^a \partial_i n^b \partial_j n^c$, $\epsilon$ is the Levi-Civita symbol, $A_i$ is defined as $F_{ij} = \frac{1}{2}(\partial_i A_j - \partial_j A_i)$, Einstein summation convention is used and details of calculation are presented in the *Methods section*. For all studied transformations, the Hopf indices obtained via integration of the topological charge density before and after re-linking of vortices match up to numerical errors, with the corresponding numerical values provided in insets of Extended Data Fig. 6. These values of Hopf index are also consistent with the geometric analysis and interpretation of this topological invariant as the linking number between preimages of any pair of distinct points on $\mathbb{S}^2$, the order-parameter space of vectorized **n(r)** (Extended Data Fig. 5)[46,47]. The Extended Data Figs. 6-8 show examples of different "isotopes" that have the same Hopf index formed from the same initial heliknoton structures but with different vortex knotting and linking details. Interestingly, fusion of elementary heliknotons helps to make complex analogues of high-baryon-number nucleons[15] or high-atomic-number chemical elements, like in the original Kelvin's vortex atom model and in topological models of nucleons[1-4].

**Elasticity-mediated interactions and reconnections of vortex knots**

The facile attractive interaction that leads to the "double reconnections" (Figs. 1 and 2) via annihilation of vortex fragments with local opposite elementary winding numbers of $\pm 1/2$ can be understood as stemming from the attractive elasticity-mediated local interaction between the vortex regions of opposite winding numbers situated in the proximity of one another. This is both analogous and different from what was observed for vortices in water[12], where the possibility of reconnections to occur depends on directionality of swirling flows around the vortex cores. At the same time, other scenarios are possible too, where locally graph-like configurations can form as transient states, embedding a superposition of reconnected states that can take place (Fig. 3a and Fig. 4). Among other notably interesting reconnection scenarios is the generation of a vortex loop that simultaneously reconnects multiple knots while serving as some kind of "glue" fusing together the dischiralation vortex cores (Fig. 4e,g and SI Video 6).

Since the Hopf index of our vortex-knot-containing structures is conserved before and after the reconnections mediating fusion and fission processes, one can ask: What is the minimum number of reconnections needed to go from one knot to another while keeping $Q$ constant? For the studied knots that all appear to have crossings of positive type both before and after fusion/fission processes, a lower bound estimate for it, and upper bound as well, can be obtained by calculating the reconnection numbers (or signature topological invariants) while following the recently introduced topological analysis of reconnections[45]. The relative reconnection number, determined as the difference $|R_A - R_B|$ between the numbers of reconnections needed to unknot each of the knots, is indeed found to be the lower bound and in some special instances equals the number of reconnections that we observe (Extended data Fig. 9). For example, the relative reconnection number for a single heliknoton undergoing internal reconnections is 2, consistent with what we observe. Furthermore, since application of special external stimuli such as very strong fields can destroy elementary heliknotons and lead to changes of Hopf index $Q$ of fused composite knot soliton states, the multiplicity of vortex reconnections associated with such knot-destroying and Q-changing transformations also need to be considered, albeit they are outside of the scope of this present study.

It is also of interest to investigate what happens with writhe during reconnections. Akin to what was found for the connected-sum-type of formation of knotted DNA molecules[42], we find that cumulative writhe is conserved during the elementary knot fusion/fission processes. However, interestingly, this is not the case during internal intra-heliknoton reconnections nor more complex reconnections that go beyond the process of fusion and fission of elementary knots (Extended Data Fig. 9).

**Chirality and topology**

Chirality of the LC host medium is essential for heliknoton stability as the chiral term in the free-energy functional allows for the (meta)stability of both the helical background and the heliknotons. Reversing handedness of the host medium gives origin to the hopfions of opposite charge $Q^{33,34}$, which can be checked by consistently vectorizing circulations of preimages[46]. The vortex knots, the preimage links and the host chiral LC medium are all chiral, with the knot chirality preserved during elementary re-linking operations within fusion and fission (Extended Data Fig. 10). In fact, all "isotopes" of vortex atoms can be thought of as obtained by re-linking operations within the knots. Switching handedness of the host medium leads to energy-minimizing knots of opposite handedness as well. Although most known knots in mathematics are chiral and achiral knots are rather rare, one can ask whether achiral knots can be possibly obtained from the fusion of elementary chiral knots. While we did not obtain such achiral knots in experiments or numerical analyses so far, we identified a scenario where specific series of band surgeries could lead to such achiral knots (Extended Data Fig. 10c), provided that both the end and intermediate states are energy-minimizing structures under suitable material parameters and voltage-driving schemes, or in response to other external stimuli. Thus, our experimentally accessible system can be used to explore the interplay between chirality at hierarchically different levels, from that of chiral centers of chiral dopant molecules within the LC to that of the chiral nematic host medium and to particle-like vortex knots embedded in it.

The topology of a chiral LC can be viewed from different perspectives. On the one hand,

the order parameter space for 3D structures of nonpolar **n(r)** is $\mathbb{S}^2/\mathbb{Z}_2$, which can be smoothly vectorized for nonsingular structures in 3D to yield an order-parameter space of $\mathbb{S}^2$. From this viewpoint, the localized field configurations are simply hopfions classified by the third homotopy groups, $\pi_3(\mathbb{S}^2/\mathbb{Z}_2)$ or $\pi_3(\mathbb{S}^2)$, which identify with the group of integers under addition, i.e., $\pi_3(\mathbb{S}^2/\mathbb{Z}_2)=\pi_3(\mathbb{S}^2)=\mathbb{Z}$. On the other hand, while considering the mutually orthogonal **χ(r)** and **n(r)** fields together, the order parameter space becomes the quotient space $\mathbb{S}^3/Q_8$,[10,48] where $Q_8$ is the quaternion group. In this framework, the vortex lines studied here can be considered as one of the elements of $Q_8$, since $\pi_1(\mathbb{S}^3/Q_8)=Q_8$, where unrestricted re-linking is allowed because of all vortex lines belonging to the same element of $Q_8$[48]. When on their own, knots of $\pi_1(\mathbb{S}^3/Q_8)$ vortex lines would not necessarily have topological protection as they could be reduced to unknots, much like in the case of vortices in water.[12] However, the dual nature of our heliknotons manifests itself in the profound conservation of a topological invariant, the Hopf index, during all observed dischiralation vortex re-linking transformations.

To conclude, we have demonstrated vortex knots in chiral liquid crystals, which exhibit a striking resemblance of vortex-knot models of atoms originally proposed by Kelvin[1]. Much like conventional atoms or nucleons, these particle-like topological objects are characterized by integer-valued Hopf index topological invariants[46,47,49], analogues of baryon and atomic numbers, and exhibit both fusion and fission, which can be electrically controlled by low-voltage fields. The ability of on-demand reconnections within arrays and crystals of knots of various symmetries by applying electric fields locally using patterned electrodes, like in displays[24], may allow for exploring combinatorial diversity of complex knots and links that can be realized within a chiral LC medium. The atom-like behaviour of our heliknotons may also allow for technological utility in electro-optics and photonics, where knots can be controlled under conditions similar to that of LC display pixels. Being realized in the spatial structures of reconfigurable optical crystals in the form of chiral LCs[19,50], our knots may induce topologically nontrivial configurations in phase or polarization of light with robust properties[14,51]. Furthermore, since similar topological objects and reconnection processes can be also realized in chiral magnets[11], the stability, fusion and fission of knots could be potentially used in spintronics, data storage and unconventional computation.

# Figures

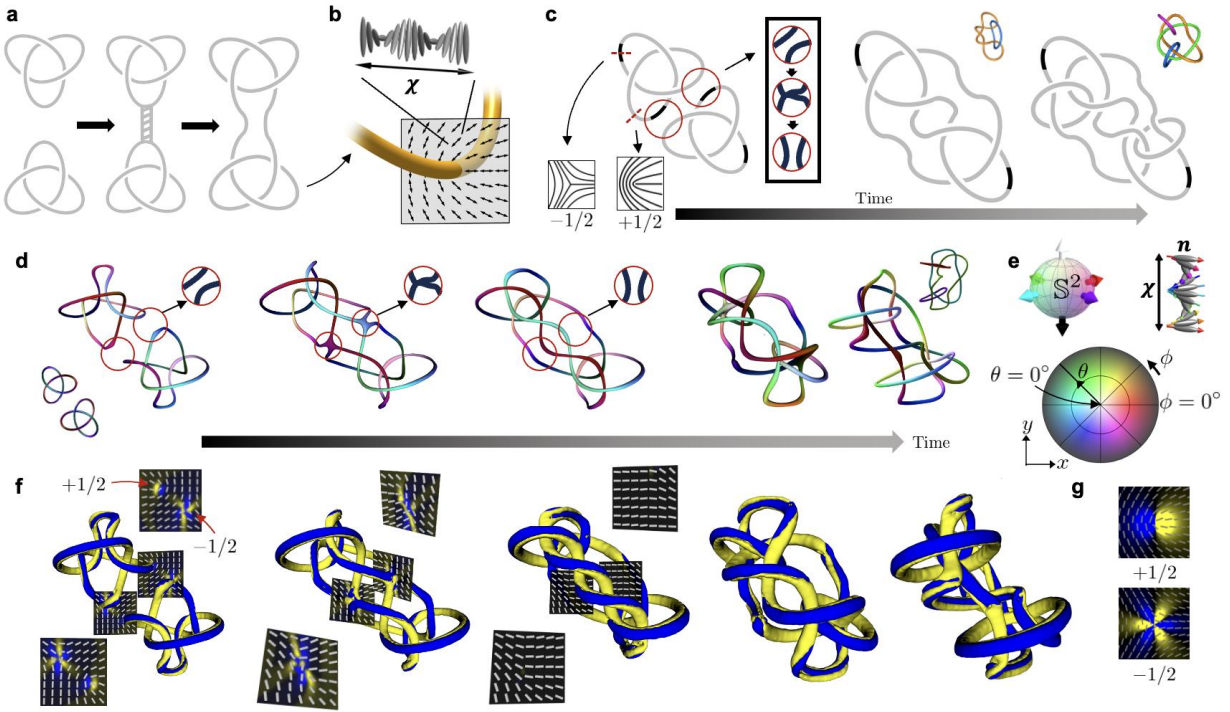

**Fig. 1 | Topological vortex reconnections in a helical twisted background of the chiral LC. a,** Connected knot sum of two trefoil knots. **b**, Schematic of a dischiralation vortex line with the core in the form of a region where the twist axis $\chi(\mathbf{r})$ is singular within a chiral LC. **c**, Schematic vortex reconnections between vortex knots of two heliknotons, where gray and black segments indicate +1/2 and -1/2 vortex line fragments, respectively. Red circles highlight regions of reconnection progressing from left to middle; additional intra-heliknoton reconnections transform dischiralation vortex knots depicted in the middle-to-right schematics (see Fig. 2i). **d**, Two heliknotons undergoing a paired reconnection event, transforming from two trefoils to a multi-component link colored according to the director orientation shown in **e**. Red circles highlight regions where reconnections progress through the intermediate formation of vertices of a four-valent graph, in a process as indicated in **c**. **e**, Color mapping scheme of vectorized director orientation based on $\mathbb{S}^2$ sphere (top left), where all possible orientations of the unit vector are uniquely represented by the colors on the unit sphere, as illustrated for the helical structure (top right). In the flattened version of the colored unit sphere (bottom), the arrows show directions of increasing azimuthal and polar angles describing orientations of $\mathbf{n}(\mathbf{r})$, where the white center corresponds to the north pole and black periphery to the south pole of the $\mathbb{S}^2$, respectively. **f**, Reconnections seen in **d** visualized with ribbons of splay and bend where dual-band and tri-band ribbons distinguish between +1/2 and -1/2 winding numbers of dischiralation lines, respectively. Positive-splay and bend regions of deformed $\chi(\mathbf{r})$ are shown in blue and negative in yellow, as depicted in **g**. Cross-sectional slices show local $\chi(\mathbf{r})$ orientation and regions of strong splay and bend. **g**, Schematic of the $\chi(\mathbf{r})$ orientation and the corresponding splay-bend geometry and color scheme for each dischiralation local structure type. Reconnections in **d, f** were initiated by reducing voltage from 2.8 V (first three frames) to 2 V (last two frames) in a cell 20 μm thick with pitch 5 μm. SI Video 1 shows the corresponding dynamics.

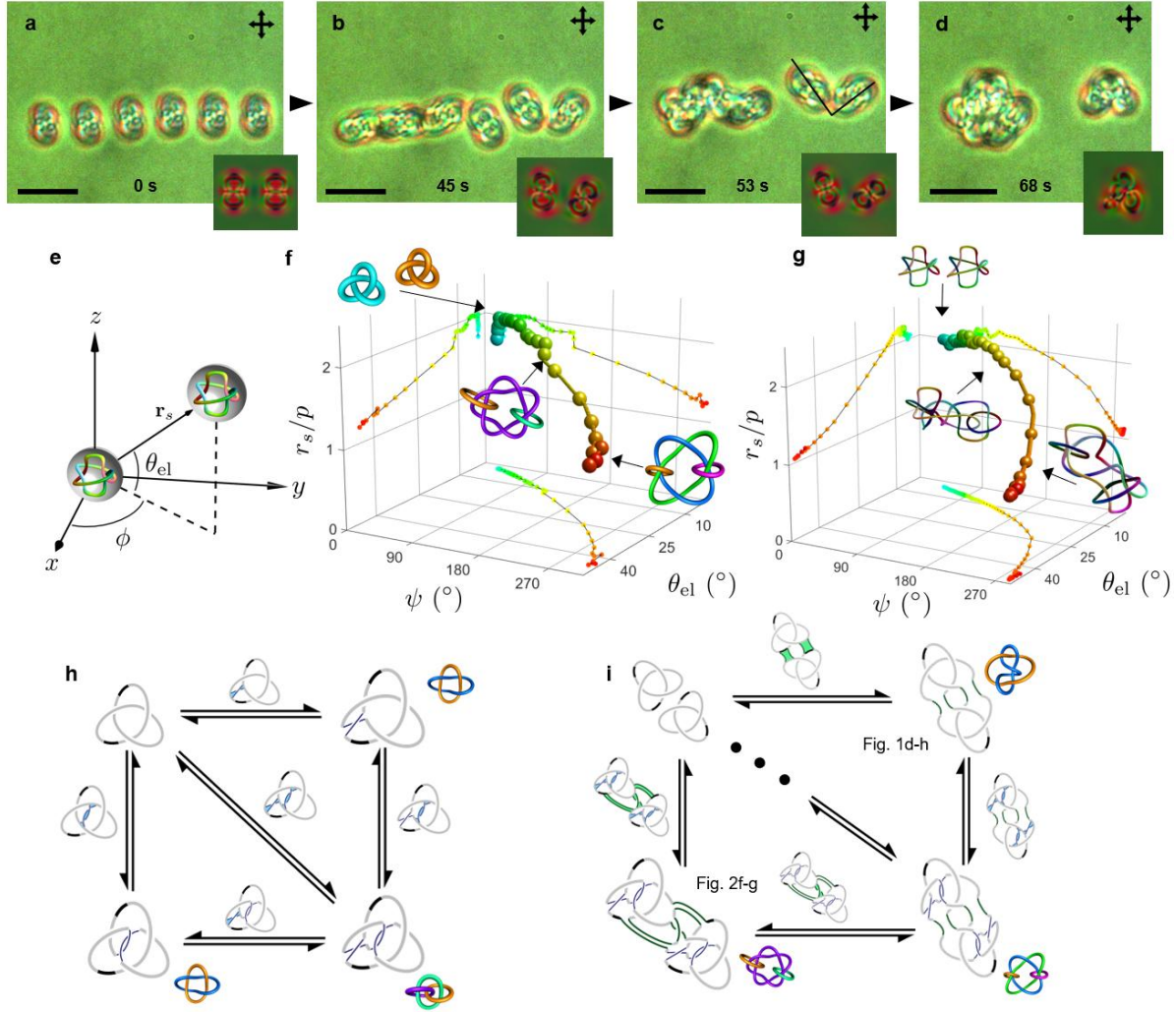

**Fig. 2 | Orientation-dependent fusion of vortex knots. a-d**, POM time series showing vortex knots after they are perturbed from the initial equilibrium configuration by changing voltage. Scale bars are 10 μm; crossed double arrows show orientation of crossed polarizers. Bottom right insets are corresponding numerically simulated POMs. Angle $\psi$ defines the relative in-plane angle between the long axes of two interacting heliknotons. $d = 16$ μm and $p = 6.9$ μm in **a-d** and $U = 1.7$ V in **a** and 2.1 V in **b-d**. SI Video 1 shows the corresponding dynamics. **e**, Relative heliknoton-heliknoton positions and the visualization depicting the orientation parameters ($\theta_{el}$, $\phi$) defined relative to the separation vector $\mathbf{r}_s$ and the uniform far-field helical background of the sample. **f-g**, Experimental (**f**) and numerically simulated (**g**) trajectories of the separation vector for the two far right trefoils in **a-d**. Knot insets show the simplified vortex topology at the beginning, middle, and end of the interaction process; see also the corresponding SI Video 3. **h**, Band surgery schematics of admissible reconnection pathways that occur within a single heliknoton. **i**, Band surgery schematics of trefoil-trefoil reconnections observed. Surgeries between heliknotons and within heliknotons are colored blue and green, respectively. **h-i**, Arrows indicate reversible pathways induced by changing the applied voltage. While different pathways can be pre-selected by means like relative initial positions, sample's thickness and pitch, surface boundary conditions, voltage amplitude, frequency and various kinetic voltage driving schemes, the detailed explorations of such means of control is beyond the present study's scope.

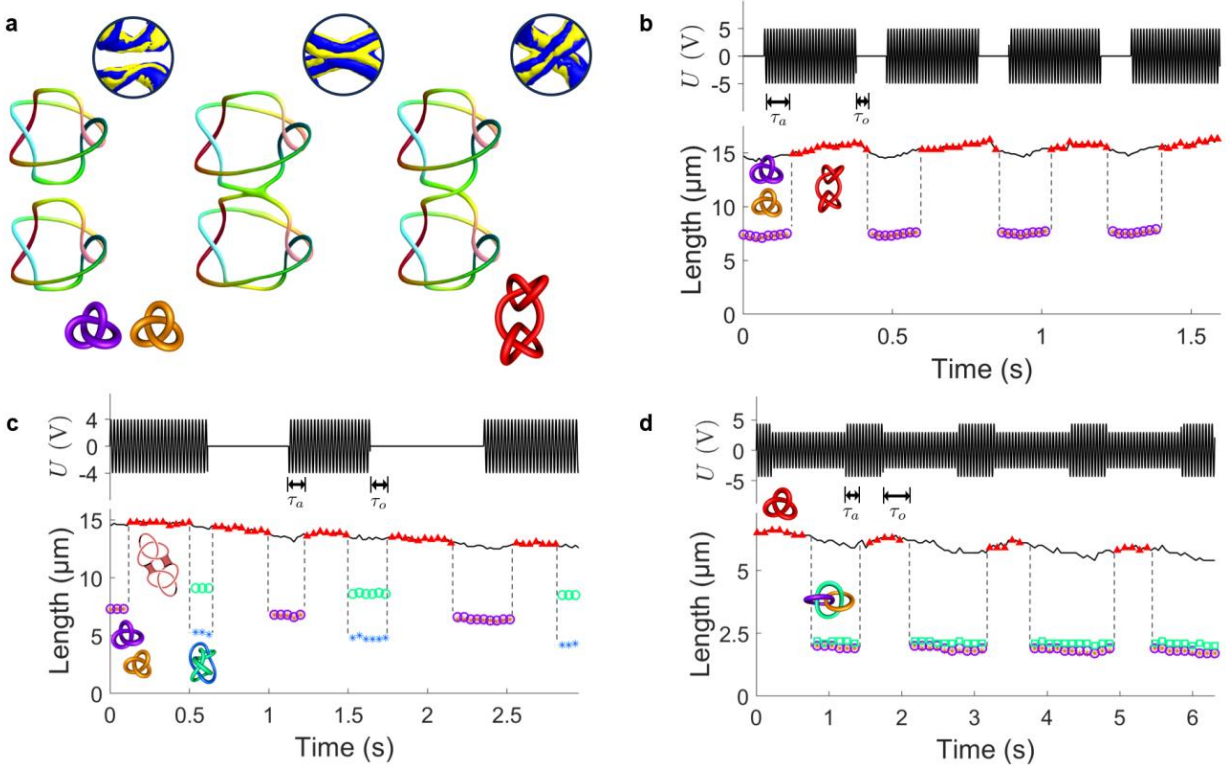

**Fig. 3 | Electrical switching of fusion and fission in vortex knots. a**, Reconnection of trefoils with top insets visualizing splay-bend ribbons at the reconnection site before (left), during (middle), and after (right) the reconnection. Top insets show the details of structure change in the region of reconnections. Bottom-inset knots show the simplified knot topology before and after the reconnection; see also the corresponding SI Video 4. **b**, Repeated fusion and fission of trefoil vortices depicted in **a**. **c**, Switching between trefoil knots and the multi-component links shown in Fig. 1d-f, repeated multiple times (SI Video 5). **d**, Switching between a single trefoil knot configuration and three linked loops. In **b-d**, Total knot length is plotted via the solid black curves; dashed black lines guide the eye when the number of components change. Top panels show the voltage magnitude of the pulse train. $\tau_a$ ($\tau_o$) denotes the time interval between a switch in electric pulses and the corresponding link-changing fusion (fission) event. The parameters used are $d = 25$ μm, $p = 5$ μm in **a-d**.

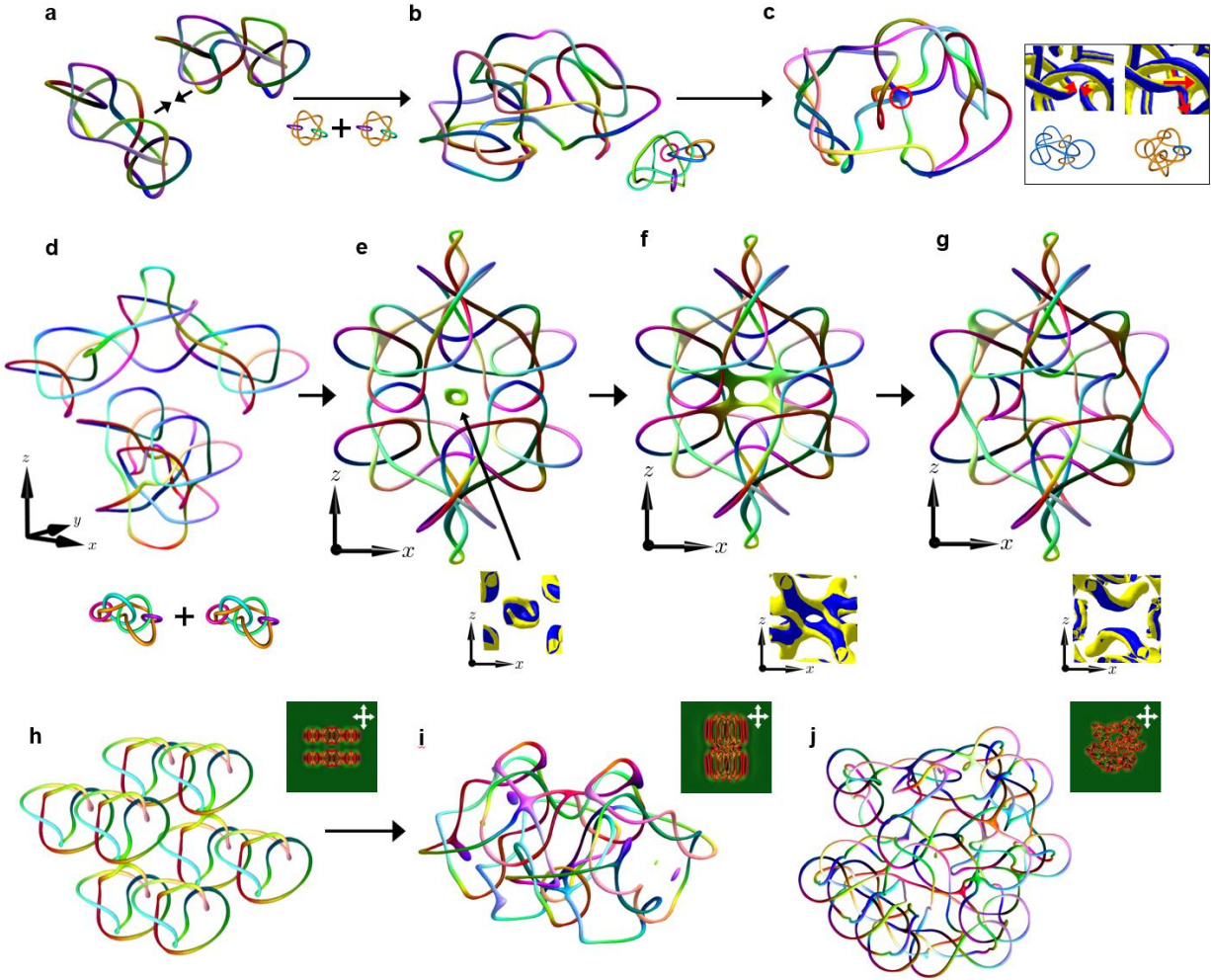

**Fig. 4 | Formation of complex knots through fusion of simple ones**. **a-b**, Two $Q = 2$ fused knot dimers, which previously formed from individual heliknotons (each stable at ~3.7V) first through fusion of pairs of individual trefoil-shaped vortex knots (Extended Data Fig. 7a-e), which then hybridize together to form a tetramer. Each stage of fusion was driven by switching voltage to ~4V. Insets show simplified vortex knot schematics. **c**, Graph of a tetramer with the circled 4-valent node that can be resolved into different states depending on relinking; the detailed configurations of vortex lines and the corresponding simplified knot schematics of the entire knots are illustrated in the boxed inset. **d-g**, Two $Q = 3$ trimers (schematically shown in the inset of **d**) reconnecting to produce a $Q = 6$ complex graph with several nodes; for dynamics, see SI Video 6. **h-i**, A fused state of eight $Q = 1$ heliknotons arranged closely together and forming a vortex graph with $Q = 8$, induced by applying 4 V (**h-i**). The detailed configurations of vortex lines in the region of fusion are shown in the bottom insets of d-g. **j**, A fused state of 18 elementary heliknotons relaxed from a perturbed lattice to form an interconnected graph with $Q = 18$ (see also Extended Data Fig. 8 and dynamics of simulated POM in SI Video 7); the complex knot is produced via fusion of individual knots from an array by pulsing (3 times) with voltage amplitudes between 1.5-3 V. Simulated POMs of the knotted structures are shown as insets in **h-j**, which are obtained for crossed polarizers indicated by white double arrows. Parameters used in simulations are $d$ = 25 μm and $p$ = 5 μm.

# Methods

## Materials and sample preparation

The chiral LC mixture used was prepared from 4-Cyano-4'-pentylbiphenyl (5CB, EM Chemicals), doped with the left-handed chiral additive cholesterol pelargonate (Sigma-Aldrich). To obtain a target pitch $p$ = 5-10 μm for the mixture, the formula $c_d = 1/(\text{HTP} \cdot p)$ where $c_d$ is the mass concentration of chiral additive and HTP = 6.25 μm$^{-1}$ refers to the helical twisting power for cholesterol pelargonate in 5CB. The resulting pitch of chiral LCs was confirmed using the Grandjean-Cano wedge cell method[10]. For samples prepared to conduct nonlinear imaging, 80% of the chiral 5CB mixture was mixed with 19% of reactive mesogen RM-257 (Merck) and 1% photoinitiator Irgacure 369 (Sigma-Aldrich). To prepare LC cells responsive to electric fields, indium-tin-oxide (ITO) substrates were spin-coated with PI-2555 (HD Micro-Systems) at 2700 rpm for 30 seconds and subsequently baked for 5 minutes at 90 °C, followed by an hour at 180 °C. The polyimide coated side was rubbed with a velvet cloth to produce a preferred planar alignment for LC molecules. The as-prepared ITO glass substrates were then assembled into LC cells using ultraviolet-curable glue with silica spheres whose diameter range from 20-30 μm inserted between the substrates to provide a well-defined cell gap. The ITO glass was then soldered with copper wires and attached to a voltage supply (GFG-8216-A, GW Instek) to control the voltage across the cell. The assembled cells were filled with chiral LC mixtures through capillary forces.

## Generating and imaging localized heliknotons in chiral LCs

Generation of heliknotons was carried out by holographic laser tweezers focusing a 10-30 mW laser beam produced by an ytterbium-doped continuous-wave fiber laser (YLR-10-1064, IPG Photonics) into the cell, when a 1 kHz voltage ~2-3 V was applied across the sample. The holographic laser tweezers setup can produce arbitrary patterns of laser intensity within the sample, though a focused beam that locally disrupts the orientational order of the LC material is generally sufficient to create initial conditions for the system that relax into heliknotons when a suitable voltage is applied. The beam only needs to be applied for a few seconds to initialize a heliknoton. Once generated, a laser power ~5 mW can be utilized to steer heliknotons and guide their interaction and assembly, thereby forming lattices, arrays, and hybridized vortex knots while simultaneously adjusting the applied voltage. The pairwise interaction between heliknotons can be modulated by increasing or reducing the voltage, initiating vortex reconnection events dependent on the positions and orientations of the heliknotons involved. Polarizing optical micrographs were taken with an IX-81 Olympus microscope incorporated with the holographic laser tweezers mentioned above, using a pair of orthogonally orientated polarizers, to allow in-situ imaging through a CCD camera (Flea FMVU-13S2C-CS, Point Gray Research). Several high numerical aperture (NA) objectives ranging from 100x, 40x, and 20x magnifications (NA = 1.4, 0.75, and 0.4, respectively) were used in experiments to observe detailed structures of individual heliknotons or assemblies and crystals of multiple heliknotons with a larger field of view.

**Three-dimensional nonlinear optical imaging**

To resolve the detailed structure within heliknotons and fused heliknoton structures, we utilized a three-photon emission fluorescence polarizing microscopy (3PEF-PM) setup which is directly integrated with the IX81 microscope described above. To prepare samples for 3PEF-PM, after generating soliton structures in a cell with the polymerizable chiral LC mixture, UV light from a 20 W mercury lamp was concentrated to a small region of interest through an aluminum foil mask with a pinhole to locally polymerize and preserve orientational order by crosslinking the reactive mesogens. The small polymerization region allows heliknotons to be generated and "frozen" at multiple spots within a cell sequentially, improving throughput. Once polymerized, the cell was split apart and most of the unpolymerized 5CB molecules were washed away with isopropyl alcohol and replaced with index-matching immersion oil. This procedure was done to minimize birefringence of the LC material, which can lead to imaging artifacts, while maintaining the LC $\mathbf{n(r)}$ configurations. A Ti-sapphire oscillator (Chameleon Ultra II, Coherent) operating at 870 nm with 140-fs pulses at 80 MHz repetition rate was used to excite the remaining 5CB molecules by three-photon absorption. The fluorescence signal is filtered with a 417/60-nm bandpass filter and detected in forward detection mode with a photomultiplier tube (H5784-20, Hamamatsu). The signal intensity of the three-photon absorption process scales with $\cos^6(\beta)$ where $\beta$ is the relative angle between the long axis of the 5CB molecule and the polarization vector of the light. For imaging scans done in this work, circularly polarized light, obtained by a quarter-wave late, was utilized to extract preimages of $\mathbf{n(r)}$ aligned along the far-field helical axis $\chi_0$, corresponding to regions with the lowest fluorescence signal. Isosurfaces extracted from experimental imaging were then analyzed and contrasted with the corresponding numerical structure relaxed from initial conditions matching the experimental starting configuration.

**Numerical modelling of heliknotons via energy minimization**

Numerical modelling of fusion and fission of heliknotons and other knotted solitonic structures is based on minimizing the Frank-Oseen free energy functional

$$F[\mathbf{n}(\mathbf{r})] = F_{\text{elastic}}[\mathbf{n}(\mathbf{r})] + F_{\text{electric}}[\mathbf{n}(\mathbf{r})] \quad (1)$$

via the finite difference method. Here $F_{\text{elastic}}$ accounts for elastic energy penalties incurred due to splay ($k_{11}$), twist ($k_{22}$), bend ($k_{33}$), and saddle-splay ($k_{24}$) deformations and takes the form

$$F_{\text{elastic}}[\boldsymbol{n}(r)] = \int d^3 r \{\frac{k_{11}}{2}(\nabla \cdot \boldsymbol{n})^2 + \frac{k_{22}}{2}[\boldsymbol{n} \cdot (\nabla \times \boldsymbol{n}) + 2\pi/p]^2 + \frac{k_{33}}{2}[\boldsymbol{n} \times (\nabla \times \boldsymbol{n})]^2 - \frac{k_{24}}{2}\nabla$$
$$\cdot [\boldsymbol{n}(\nabla \cdot \boldsymbol{n}) + \boldsymbol{n} \times (\nabla \times \boldsymbol{n})]\}. \quad (2)$$

The saddle-splay energy contributes to energetics of defects and surface anchoring energy[52,53]. Since the **n(r)** configurations considered in this work are continuous and strong anchoring conditions are applied, we set $k_{24}$ to zero. The other elastic constants $k_{11}$ (6.4 pN), $k_{22}$ (3.0 pN), and $k_{33}$ (10.0 pN) take the experimentally determined values of 5CB. Similarly, the electric contribution is defined by

$$F_{\text{electric}}[\boldsymbol{n}(r)] = -\frac{1}{2}\varepsilon_0 \int d^3 r [\varepsilon_\perp E^2 + \Delta\varepsilon(\boldsymbol{n} \cdot \boldsymbol{E})^2], \quad (3)$$

where $\boldsymbol{E}$ is the applied field, $\varepsilon_0 = 8.854 \cdot 10^{-12}$ F/m is the permittivity of the vacuum, $\varepsilon_\perp$ is the dielectric coefficient perpendicular to the director axis, and $\Delta\varepsilon$ is the dielectric anisotropy of the LC medium. For 5CB, $\Delta\varepsilon$ and $\varepsilon_\perp$ take the values 13.8 and 5.2, respectively. The equations of motion for the director field were obtained from varying the total energy functional and replacing derivatives with their fourth-order finite difference counterparts. This subsequently, yields a set of coupled algebraic equations at each grid point to locally update the director field. To ensure numerical stability, an underrelaxation routine was performed such that the successive numerical solution is a weighted average of the old and new solution, $n_i \rightarrow \alpha n_i + (1 - \alpha)n_i'$. The parameter $\alpha \in [0,1]$ is generally set to $\alpha = 0.1$ and was chosen empirically to ensure a convergent solution. To account for local distortions in the electric field due to the dielectric nature of the LC material, after each subsequent update for the director field, the voltage is updated by minimizing the free energy with respect to the electric field and substituting derivatives of the voltage with fourth-order finite difference derivatives. This yields an equation to update the electric field at each grid point evolving the voltage simultaneously as the director field is relaxed. Periodic boundaries were assigned to the planes perpendicular to the helical axis, while hard boundary conditions were defined on the top and bottom of the cell. High throughput grid calculations were performed in parallel via code written in C++ with CUDA acceleration.

Heliknotons are initialized from the ansatz[6,9,17]

$$\mathbf{n}'(r) = q(r)^{-1}\vec{N}_{bg}(r)q(r) \quad (4)$$

where $\vec{N}_{bg}(\vec{r}) = \cos(2\pi z/p)\hat{x} + \sin(2\pi z/p)\hat{y}$ defines the background helical director field, $p$ is the pitch, and $q(r) = \cos(\pi Q r/p) + \sin(\pi Q r/p)\hat{r}$ is a quaternion. Here, $Q$ is an integer that defines the charge of the heliknoton and is set to unity for elementary heliknotons. In our modeling, to obtain the final heliknoton ansatz $n(r)$, the z-component of $n(r)$ is inverted:

$$n_x = n'_x, \quad n_y = n'_y, \quad n_z = -n'_z. \quad (5)$$

For initial conditions involving multiple elementary heliknotons, a cutoff radius is chosen to be the pitch $p$ to allow for the embedding of multiple heliknotons in a uniform helical background. This is carried out by superimposing the ansatz configurations for heliknotons localized at different locations $\{r_i\}$. The ansatz above is then relaxed according to the energy minimizing procedure described above.

To calibrate the elapsed time in simulations to match that of experiments, we reconstruct the same initial conditions for both experiment and numerical simulation of the reconnection corresponding to Fig. 2 and Extended Data Fig. 7. For a node density of $25^3 \, p^{-3}$, we find the time elapsed for each iteration to be 0.205 ms.

**Fusion and fission response times**
We explore the dynamic characteristics of elementary heliknotons fusing and splitting apart by pulsing the applied voltage. Here we describe the dependence of the response times $\tau_a$ ($\tau_o$) as a function of the voltage amplitude in detail. Although a complete characterization of the switching dynamics is beyond the scope of this work, the most relevant parameter allowing one to tune switching dynamics is the applied voltage magnitude (see Extended Data Fig. 7 f,g). Given a pair of heliknotons in the trefoil state just prior to (after) a fusion event, the response time $\tau_a$ ($\tau_o$) diverges for a critical voltage dependent on the separation distance and relative orientation of the heliknoton-heliknoton pair. This critical voltage can be interpreted as the voltage that typically stabilizes a four-valent intermediate graph topology instead of completing the reconnection event. As the applied voltage deviates from the critical voltage, the response times sharply drop and saturate to finite sub-second values. Note the response times in (Extended Data Fig. 7 f,g) are somewhat smaller than those observed in Fig. 3b,c in the main text due to the closer distance at which the heliknotons were initialized in this context. The set of parameters used above can be translated to other experimental geometries by noting the Frank-Oseen free energy functional can be rescaled by the pitch without influencing the equations of motion. Thus, the response time (neglecting initial conditions) appears to depend only on the dimensionless electric field defined by $\tilde{E} = \sqrt{\varepsilon_0 \Delta\varepsilon/\overline{K}}(V/\tilde{d})$ where $\tilde{d}$ is the thickness of the cell expressed in units of the pitch $p$ and $\overline{K}$ is the average elastic constant.

**Visualization and topological characterization of heliknotons**

The helical-axis field, $\boldsymbol{\chi}(\mathbf{r})$ is obtained by constructing the chirality tensor $C_{ij} = \partial_i n_l \epsilon_{jlk} n_k$ where Einstein summation convention is assumed and obtaining the dominant principal eigenvector which defines the orientation of the local nonpolar helical axis $\boldsymbol{\chi}(\mathbf{r})$. Local regions within heliknotons that do not have a well-defined chiral axis correspond to vortex lines. In this work, we color these vortex lines according to their local director orientation on the $\mathbb{S}^2$ sphere. Vortex knots obtained by sampling the raw grid points are often coarse. To improve the quality of these knots, vortices are first smoothed via Taubin smoothing to ensure a faithful reconstruction of the knot topologies[54]. This smoothed isosurface data is then used to construct a graph, which is traversed to find link components before and after knot reconnections. When graphs can be successfully resolved into links, the corresponding knot diagrams are imported into KnotPlot where they can be relaxed and further analyzed[55]. Ribbons of splay-bend in a tubular neighborhood about the vortex lines are produced by constructing a tensor $\mathbb{Q}(\boldsymbol{\chi}) = \boldsymbol{\chi} \otimes \boldsymbol{\chi} - 1/3$ and calculating the splay-bend parameter $S_{\text{SB}} = \partial_i \partial_j \mathbb{Q}_{ij}$ (here Einstein summation is assumed)[56]. The blue and yellow ribbons indicate regions with $S_{\text{SB}} > 0$ and $S_{\text{SB}} < 0$, respectively, and correspond to isosurfaces of $S_{\text{SB}}$ values 10% of the average positive splay-bend $\langle S_{\text{SB}}^+ \rangle$ and 10% of the average negative splay-bend $\langle S_{\text{SB}}^- \rangle$ within the tubular neighborhood of the vortex knot. To produce smooth ribbons close to vortex cores where $\boldsymbol{\chi}$ is ill-defined, $S_{\text{SB}}$ at each grid point is locally averaged with its nearest neighbors.

Hopf indices of elementary and hybridized heliknotons are computed numerically according to the following procedure described elsewhere[38,47]. First, we make the identification $b^i = \epsilon^{ijk} F_{jk} = \epsilon^{ijk} \partial_j A_k$ allowing us to associate the quantity $\boldsymbol{A}$ with a vector potential of $\boldsymbol{b} = \boldsymbol{\nabla} \times \boldsymbol{A}$. The Hopf index $Q$ can then be written as $Q = \frac{1}{64\pi^2} \int d^3 r \, \boldsymbol{b} \cdot \boldsymbol{A}$. It follows that upon computing $\boldsymbol{b}$, the vector potential $\boldsymbol{A}$ is obtained from numerical integration and the Hopf index $Q$ can be obtained. All numerical derivatives are performed with fourth-order accuracy yielding Hopf indices that agree within numerical error with the number of heliknotons initialized. The Hopf charge may also be determined by counting the linking number of different vectorized preimages[46]. The north- and south-pole preimages corresponding to director orientations along $\boldsymbol{\chi}_0$, or the z-axis. These preimages can be numerically extracted by computing isosurfaces according to the condition $|\mathbf{n}(\mathbf{r}) - \mathbf{n}_t| < \eta$ where $\eta$ is a numerical tolerance set to 0.1 corresponding to a small neighborhood of allowed vectorized $\mathbf{n}(\mathbf{r})$ orientations surrounding the target orientation $\mathbf{n}_t$.

Simulated POM movies were generated by applying a simple Jones matrix approach. We begin with a homogeneous input vector $\mathbf{E}_0 = (1 \quad 0)^T$ representing linearly polarized light along the *x*-axis of a given wavelength λ. Rays of light are assumed to propagate along the far-field helical axis $\chi_0$ (along the *z*-axis) of the cell followed by a crossed polarizer aligned with the *y*-axis. For a small LC volume of thickness Δ*z* with the director aligned with the *x*-axis, the corresponding Jones matrix is

$$J_0 = \begin{pmatrix} e^{i\delta_{\text{eff}}} & 0 \\ 0 & e^{i\delta_0} \end{pmatrix} \quad (6)$$

where $\delta_0 = 2\pi n_0 \Delta z/\lambda$ and $\delta_{\text{eff}} = 2\pi n_{\text{eff}} \Delta z/\lambda$ are the phases of the fast and slow axes, respectively. The extraordinary ($n_e$) and ordinary ($n_o$) refractive indices are related to the effective refractive index accounting for the out-of-plane angle θ of the director and is given by

$$n_{\text{eff}} = \frac{n_o n_e}{\sqrt{\cos^2(\theta) n_e^2 + \sin^2(\theta) n_0^2}} \quad (7).$$

In a medium of 5CB, $n_e$ and $n_o$ assume the values of 1.77 and 1.58, respectively. More generally, for directors with an angle $\varphi$ from the *x*-axis in the *xy*-plane, a rotation $R(\varphi) \in SO(2)$ can be applied to $J_0$ according to $J(\theta, \phi) = R(\varphi) J_0(\theta) R(\varphi)^T$. Applying this Jones matrix ansatz to the discretized grid geometry above, the effective Jones matrix for each point (*x*,*y*) in the focal plane is obtained by multiplying successive Jones matrices from different layers together corresponding to a column with $N_z$ elements along the helical axis:

$$M(x, y) = \prod_{1 \leq i \leq N_z} J\big(\theta(x, y, z_i), \varphi(x, y, z_i)\big). \quad (8)$$

The output polarization for a given wavelength is obtained from applying $M(x,y)$ to the input polarization and selecting the second component $E_y^{(\lambda)}$ due to the output polarizer. The normalized intensity is computed from the squared magnitude of the output. This procedure is carried out for 650 nm, 550 nm, and 450 nm light, with relative intensities 1.0, 0.6, and 0.3, respectively, determined by the spectral content of the light source used in experiments. For still-POMs (Extended Data Fig. 5), the open-source software Nemaktis[57] with the ability to model more complex optical effects via ray-tracing and beam propagation (Extended Data Fig. 5b, bottom panel) was found yielding images generally consistent with the ones modelled by the Jones matrix approach. We found that both our Jones matrix approach and Nemaktis yield results that agree well with experiments.

**Tracking interactions between heliknotons via POM imaging**

To track the separation vector between two heliknotons during fusion and fission using POM, we make use of their key property: heliknotons have orientations and positions along the far-field helical axis coupled, thus undergoing a screw-like rotational motion when translated along the far-field helical axis[10]. In the POM video (see SI Video 2), by recording the change of a relative angle describing heliknoton's azimuthal orientation, the heliknoton's dynamics across the sample thickness (along the *z*-axis and far-field helical axis) can be tracked, in addition to tracking its lateral displacement. It follows that by defining $\psi$ to be the relative angle between the long axes of the two heliknotons, one obtains $\psi = 2\pi s_z/p$, where $s_z$ is their separation in *z*. Since the in-plane heliknoton separation can be determined from POM images directly, the full separation vector between the two heliknotons can be reconstructed. The same procedure can be applied to simulated POM images of numerically simulated heliknotons as well, to enable a direct comparison of fusion/fission between experiments and simulations. In experimental cells that are less than $4p$ in thickness, heliknotons tend to persist in the mid-plane of the cell allowing $\psi$ to be easily determined as the heliknotons are perturbed from equilibrium by changing the voltage or laser tweezer manipulation.

**Characterization of knot topology**

A crucial aspect of our findings is the translation of our vortex knots, and the simplified diagrams introduced to faithfully represent their topologies and site-specific reconnections. We choose to represent these reconnections in diagrams via blue and green bands corresponding to band surgeries associated with internal and external heliknoton reconnections, respectively (Fig. 2h,i and Extended Data Fig. 9d-g). Additionally, information about the local winding number of the vortices is also important as reconnections often occur through a reconnection mechanism involving the annihilation (fusion) or pair creation (fission) between vortex segments of opposite winding number. We find that for all links obtained, all reconnections analyzed can be identified with the mathematical operation of coherent band surgery where orientations are preserved.[3,45] From the diagrams produced, one can track the evolution of the writhe as vortex knots and links undergo reconnections.[3,42-45] The writhe serves as a simple measure of complexity in the knots we obtain as they are generated from right-handed trefoil building blocks where the action of incorporating another trefoil into a complex composite knot only increases the writhe (Extended Data Fig. 9).

Like the writhe, one can compute the so-called reconnection number for a given knot or link. The reconnection number of a knot or link is the least number of reconnections that need to be performed to transform it to an unknot.[43,45] In general, this number is not known, but computable bounds on it from below exist (such as the so-called signature of the knot) and a very particular upper bound is always known that we shall call the *R*-number of the knot or link and denote by *R(K)*, where *K* is the link. In this case, one simply smooths crossings such that the local knot orientation is preserved (Extended Data Fig. 9a).[45] The circles generated by this action are called Seifert circles.[3,45] The R-number is defined as[45]

$$R = c - s + 1, \qquad (9)$$

where $c$ is the number of crossings in the original diagram and $s$ is the number of Seifert circles. The meaning of the formula is that one can perform reconnections at $R$ many crossings (it is less than the total number of crossings) and obtain an unknot.[45] This is shown in Extended Data Fig. 9b for the trefoil knot and implicitly for other examples in the figure. Once one has the reconnection numbers for a reconnection pathway ($A \rightarrow B$), it follows that the relative *R*-number $|R_A - R_B|$ is a well-defined quantity that estimates from above the minimal number of reconnections necessary to transform $A$ to $B$. If the link $K$ has all positive crossings (as in Extended Data Figure 9) then $R(K)$ is equal to the reconnection number of $K$. In general, for any $K$, the link or knot $K$ can be transformed to an unknot in $R(K)$ reconnections. $R(K)$ is least among all possible unknottings when $K$ is positive.

**Methods References:**

**Acknowledgements:** We thank T. Lee and H. Zhao for discussions and technical assistance.

**Funding:** This research was supported by the U.S. Department of Energy, Office of Basic Energy Sciences, Division of Materials Sciences and Engineering, under contract DE-SC0019293 with the University of Colorado at Boulder.

**Author Contributions:** D.H. and J.-S.B.T. performed experiments and numerical modelling, under the supervision of I.I.S. D.H., L.H.K., J.-S.B.T. and I.I.S. analyzed topological invariants of knots. I.I.S. conceived and directed the research. D.H. and J.-S.B.T. and I.I.S. wrote the manuscript, with input from all authors.

**Additional Information:** Supplementary Information is available for this paper.

This Supplementary Information includes Supplementary Videos 1-10.

**Extended Data Figures and Captions**

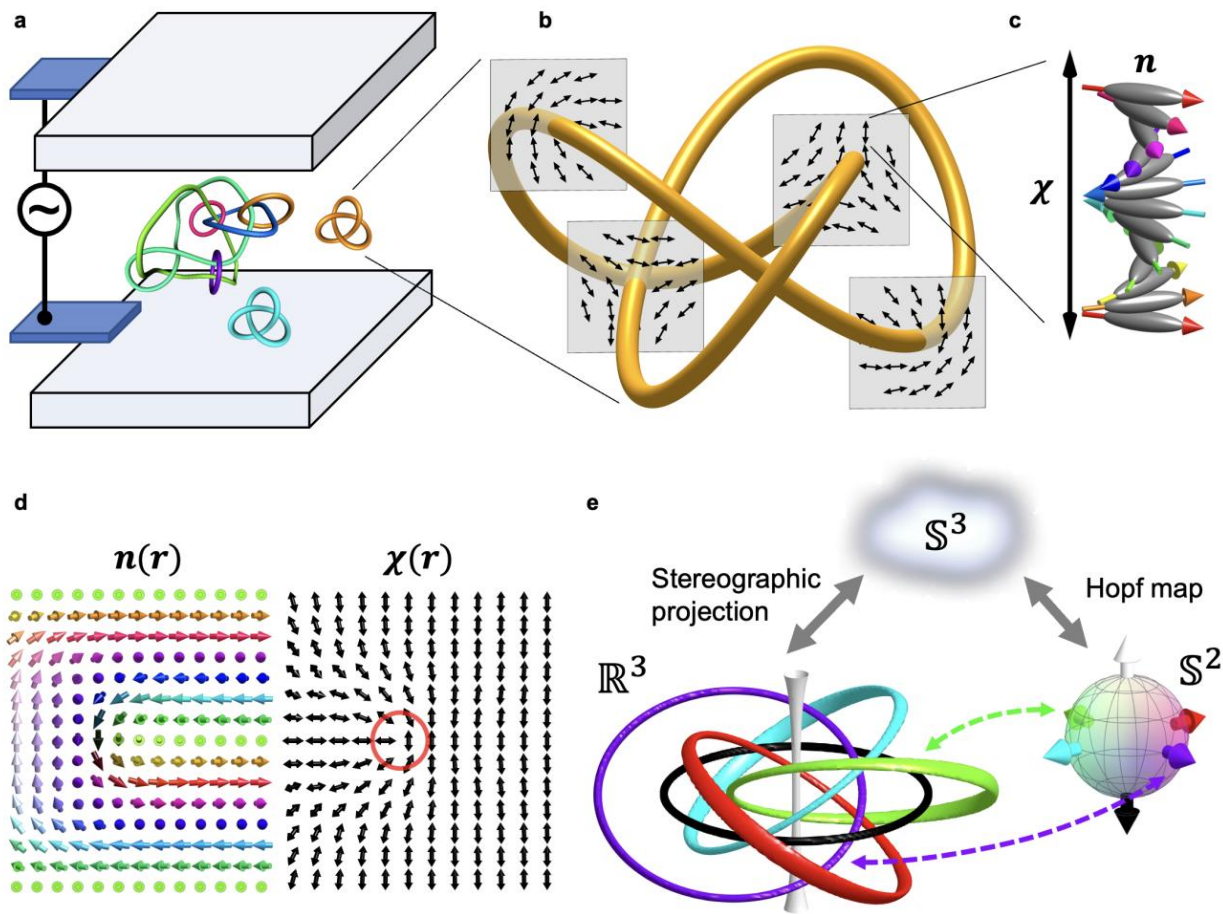

**Extended Data Fig. 1 | Heliknotons as both dichiralation vortex knots and hopfions**. **a**, LC cell geometry with ITO coated substrates, allowing to apply tunable voltage. **b**, Schematic vortex knot with $\chi(\mathbf{r})$ cross sections depicting local $\chi(\mathbf{r})$ field around the vortex tube. **c**, Schematic showing that $\chi(\mathbf{r})$ is the helical-axis field around which the LC molecules and director field $\mathbf{n}(\mathbf{r})$ twist. **d**, Twist in the director field $\mathbf{n}(\mathbf{r})$ in the cross-section of heliknoton (left) and the corresponding helical-axis field $\chi(\mathbf{r})$ (right). The red circle indicates a -1/2 dischiralation region in the vortex knot's cross-section. **e**, Hopfion topology of the heliknoton in $\mathbf{n}(\mathbf{r})$: Preimages in $\mathbb{R}^3$ (and $\mathbb{S}^3$) correspond to distinct points in $\mathbb{S}^2$ form interlinked closed loops.

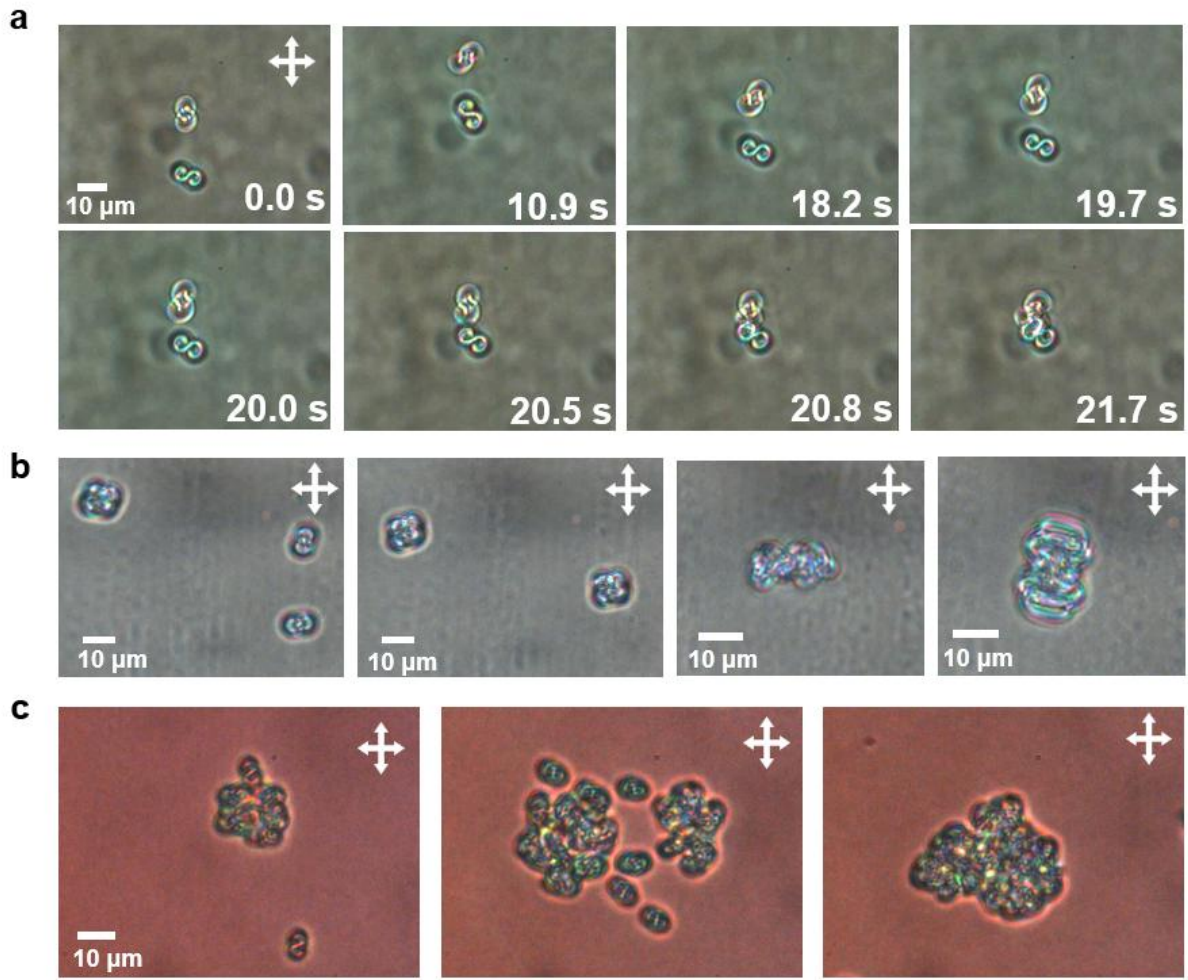

**Extended Data Fig. 2 | POM micrographs of laser-tweezer-driven vortex reconnections**. **a**, Two elementary heliknotons spontaneously fusing into a dimer; here $d$ = 30 µm, $p$ = 5 µm, and $U$ = 3.4 V. **b, c**, Laser tweezer manipulation of heliknotons to construct more complex vortex knots by incrementally fusing elementary ones greater complexity (**b**) and obtaining a "tangle" of fused heliknotons (**c**). Crossed polarizer orientations are indicated by white double arrows. The relevant parameters are $d$ = 15 µm, $p$ = 4.5 µm, and $U$ = 2.1 V in **b** and $d$ = 17.5 µm, $p$ = 5.4 µm, and $U$ = 1.8 V in **c**.

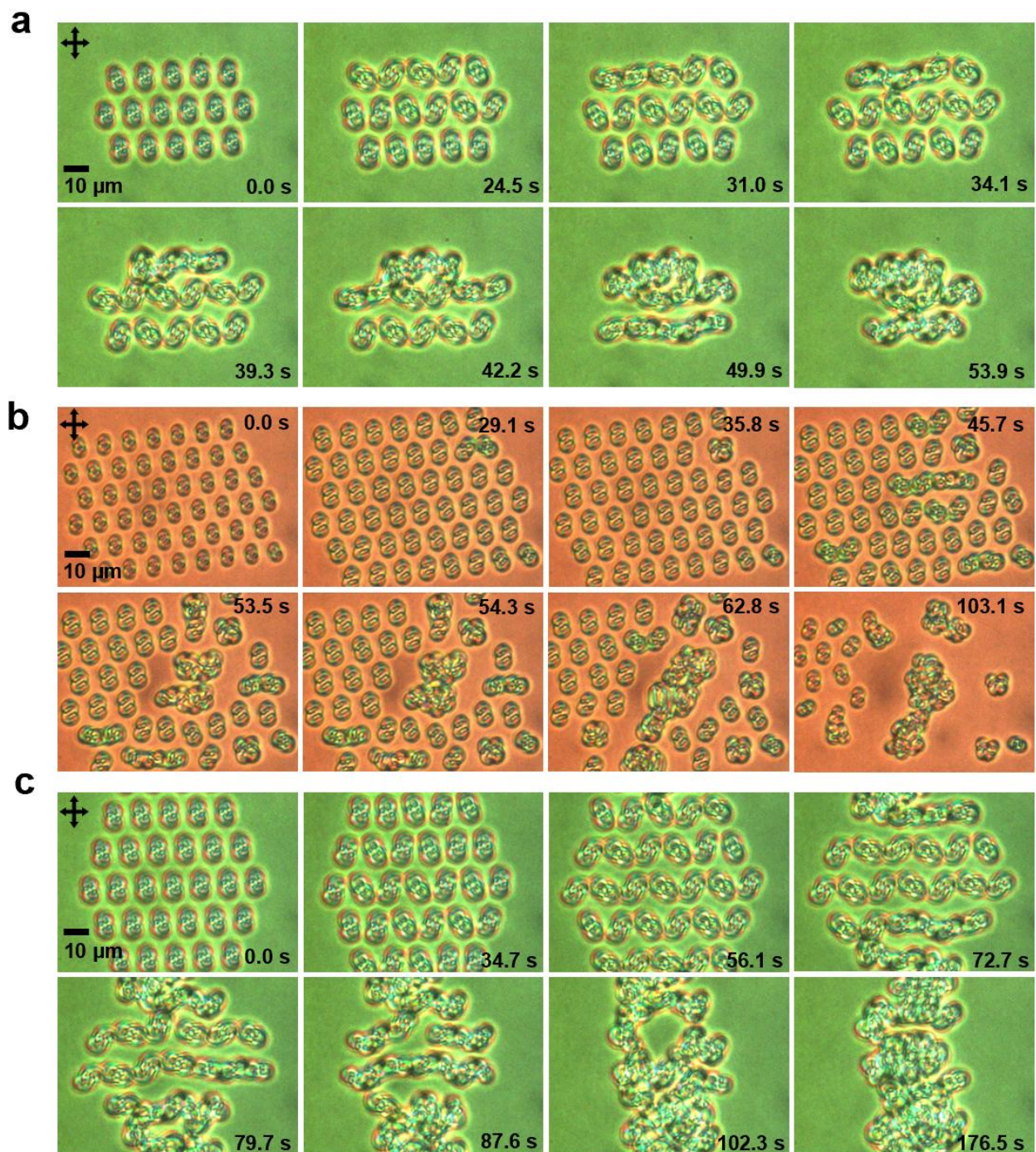

**Extended Data Fig. 3 | Fusion of complex knots from a lattice of elementary heliknotons. a-c**, Time evolution of several heliknoton lattices perturbed from equilibrium by increasing the applied voltage. Black double arrows indicate crossed polarizer orientations. In (**b**), $d$ = 17.5 μm, $p$ = 5.4 μm, and $U$ = 1.8 V in the first frame and 2.3 V in subsequent frames. In (**a**) and (**c**), $d$ = 16 μm, $p$ = 6.9 μm, and $U$ = 1.7 V in the first frame and 2.1 V in subsequent frames. The real-time dynamics of transformations corresponding to **a** and **c** is shown in the SI Video 8.

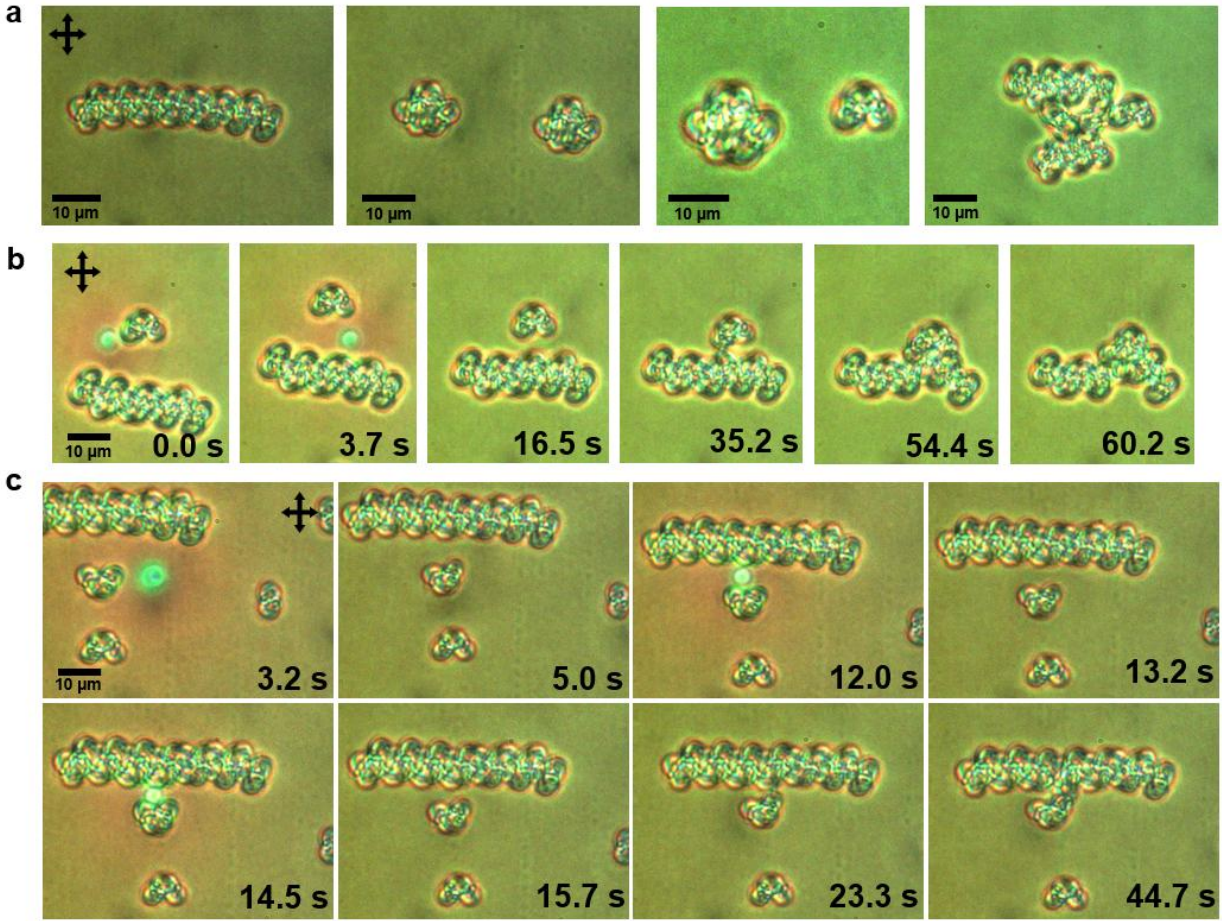

**Extended Data Fig. 4 | Laser-tweezer-guided fusion of chains and clusters of heliknotons. a**, Various fused heliknoton assembly guided by laser tweezers. **b-c**, In-situ optical manipulation of heliknoton chains fusing with heliknoton dimers at different controlled contact sites. The corresponding real-time dynamics is shown in the SI Video 9. Black arrows show crossed polarizer orientations. In (**a-c**), $d$ = 16 μm, $p$ = 6.9 μm, and $U$ = 1.7 V.

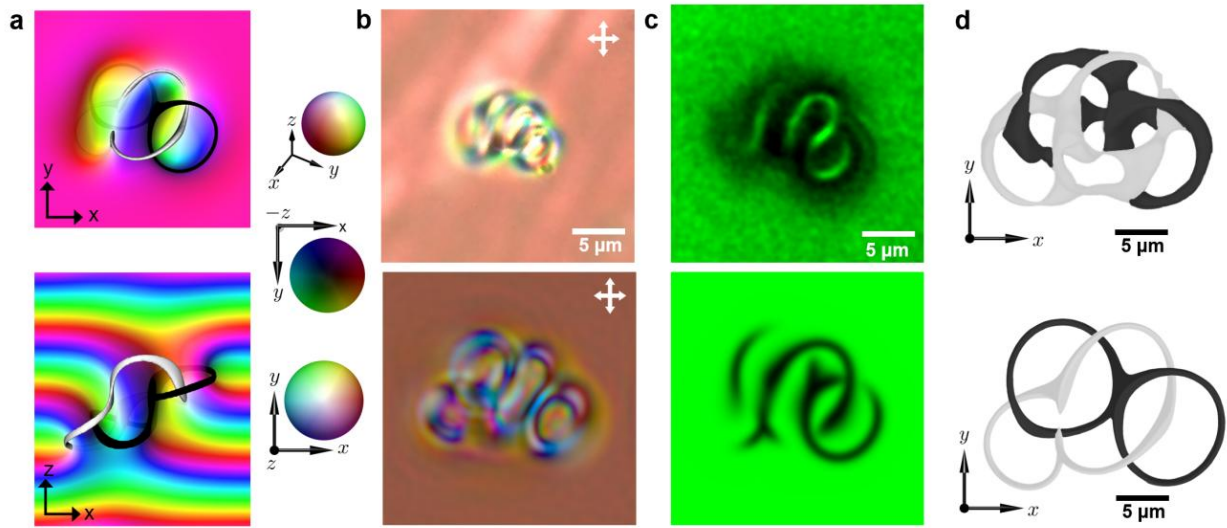

**Extended Data Fig. 5 | Structure of fused heliknotons reconstructed numerically and experimentally. a**, Simulated midplane cross-sections of a fused heliknoton. Cell thickness 20 um, pitch 5 um, applied voltage 2.8 V. White and black loops visualize the north- and south-pole preimages of vectorized director, respectively. **b**, Experimental (top) and numerical (bottom) POM images of the fused heliknotons shown in **a**. White arrows indicate crossed polarizer orientations. **c**, Reconstructed experimental (top) and numerical (bottom) nonlinear fluorescence images using obtained circularly polarized laser excitation. **d**, Polar preimages extracted from experimental 3PEF-PM imaging (top) and corresponding numerical simulations (bottom).

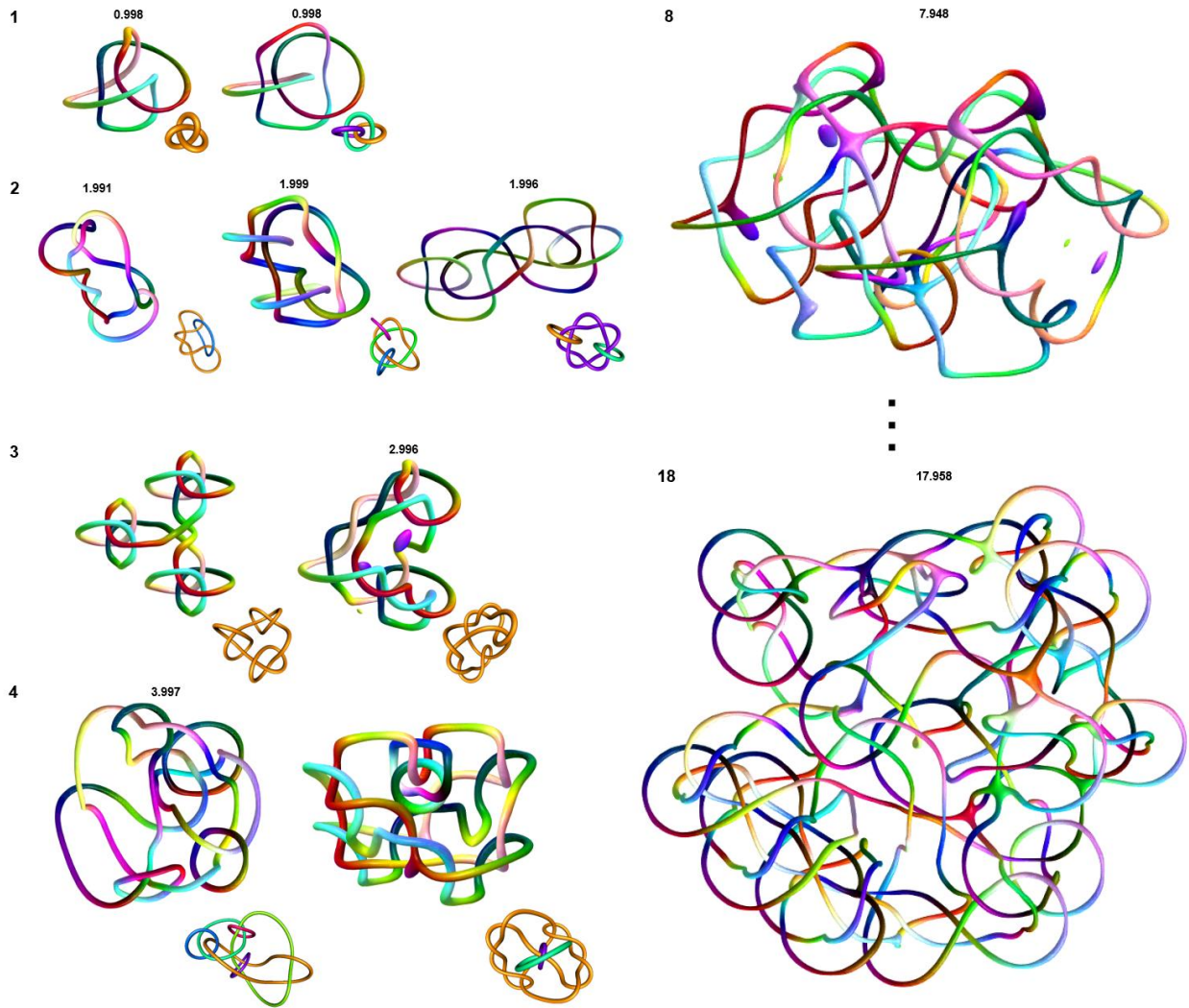

**Extended Data Fig. 6 | Zoo of knots with different Hopf indices obtained via knot fusion.** Integers at the top-left of each row are the expected Hopf index and numbers above each structure are the corresponding numerically computed Hopf indices. SI Video 10 shows dynamics of fusion of heliknotons with the net total Hopf index of $Q = 3$ and $Q = 4$. Bottom right insets are the simplified multi-component links generated by KnotPlot software.

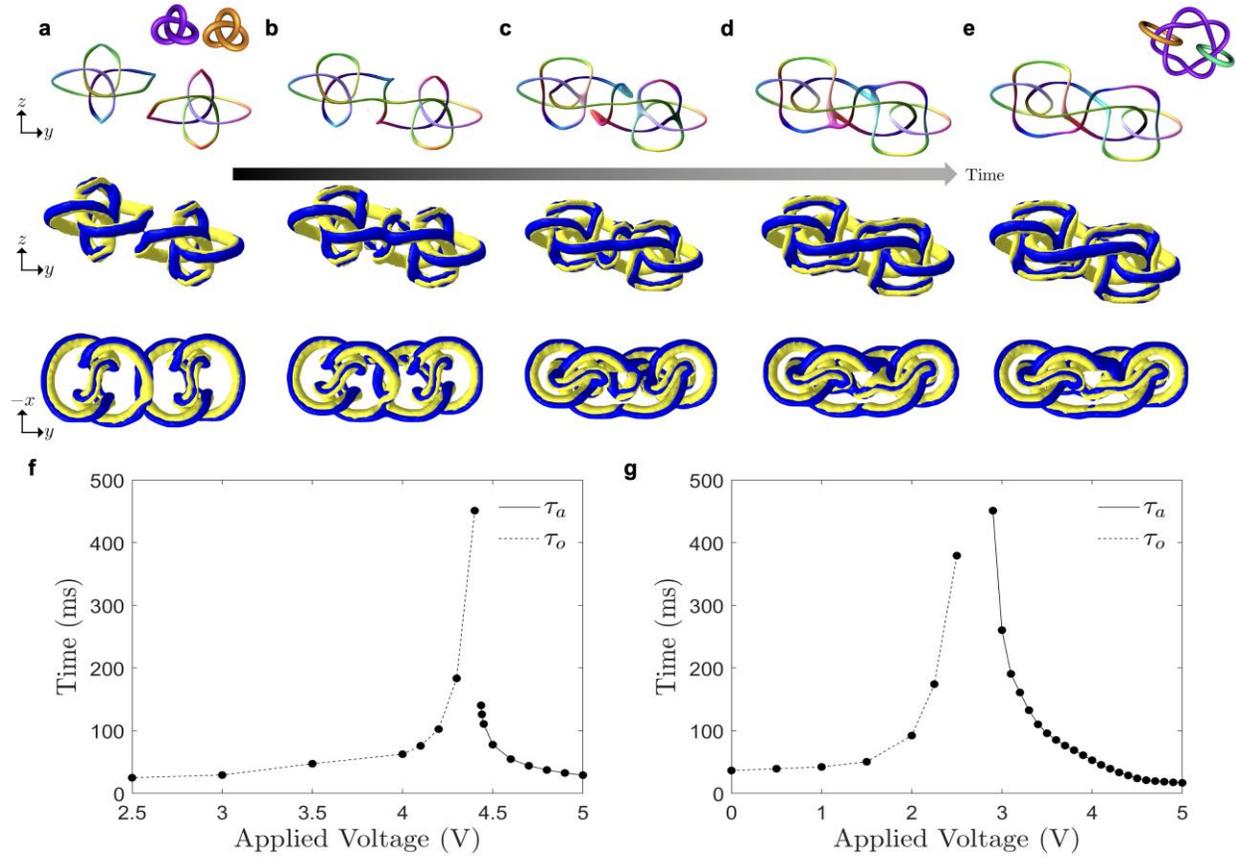

**Extended Data Fig. 7 | Vortex reconnections during fusion of two heliknotons. a-e**, Vortex reconnections from two separate heliknotons into a three-component link visualized with colored vortex knots and ribbons of splay and bend. The parameters used are: $d = 25$ μm, $p = 5$ μm, and $U = 3.9$ V. **f,g,** Response times for two heliknotons reconnecting along the far-field helical axis $\chi_0$ (as shown in Fig. 3a-b) and two heliknotons reconnecting while approaching each other at 45 degrees with respect to $\chi_0$ (corresponding to Fig. 1d,f and Fig. 3c). Simulations were performed in a cell with $d = 25$ μm and $p = 5$ μm.

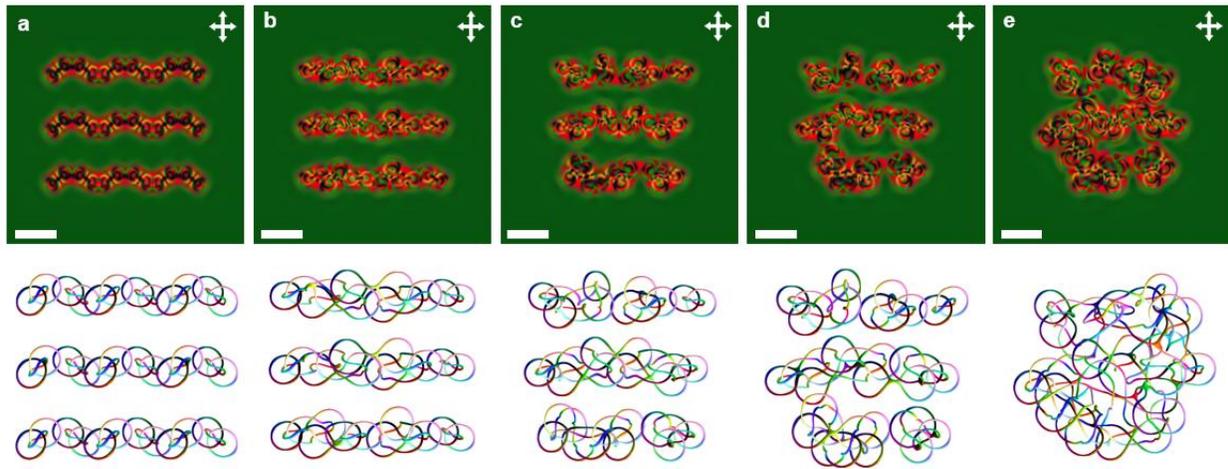

**Extended Data Fig. 8 | Numerically simulated heliknoton lattice hybridization with $Q$ = 18. a-e**, Evolution of a heliknoton lattice (**a**) pushed from an initial configuration into a knotted graph (**e**) by pulsing with voltages between 1.5-3 V (**d-e**) in a cell with thickness 50 μm and pitch 10 μm. The simulation was performed in the one-constant approximation to reduce computation time.

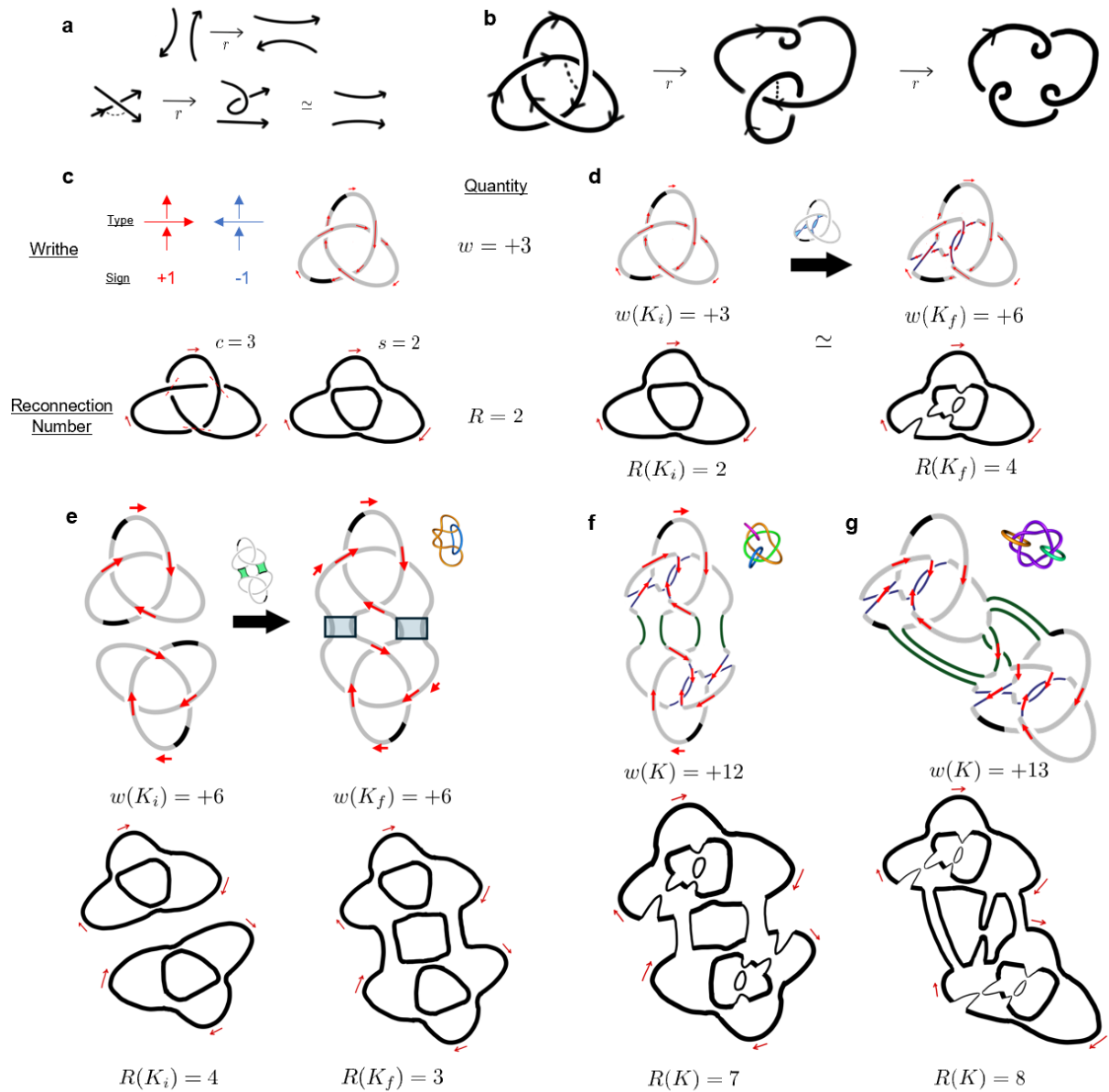

**Extended Data Fig. 9 | Writhe and reconnection numbers for knot transformations. a,** A reconnection event between two vortices (top). A reconnection event ($r$) at a crossing results in a smoothed crossing (bottom). **b,** Reconnection of a trefoil knot. **c,** Calculations for writhe (top) and reconnection number (bottom) for a trefoil knot. Red arrows serve as guides to the eye to calculate the local orientation of the trefoil knot. Here $c$ refers to the number of crossings and $s$ to the number of Seifert circles obtained after smoothing the crossings. **d, e,** Writhe and reconnection numbers before and after reconnection for a single trefoil (**c**) and two trefoil knots (**e**). **f, g,** Oriented knot diagrams and their corresponding Seifert circle diagrams used to compute the writhe and reconnection numbers. **e-g,** Red arrows marking crossings indicate a positive contribution to the writhe.

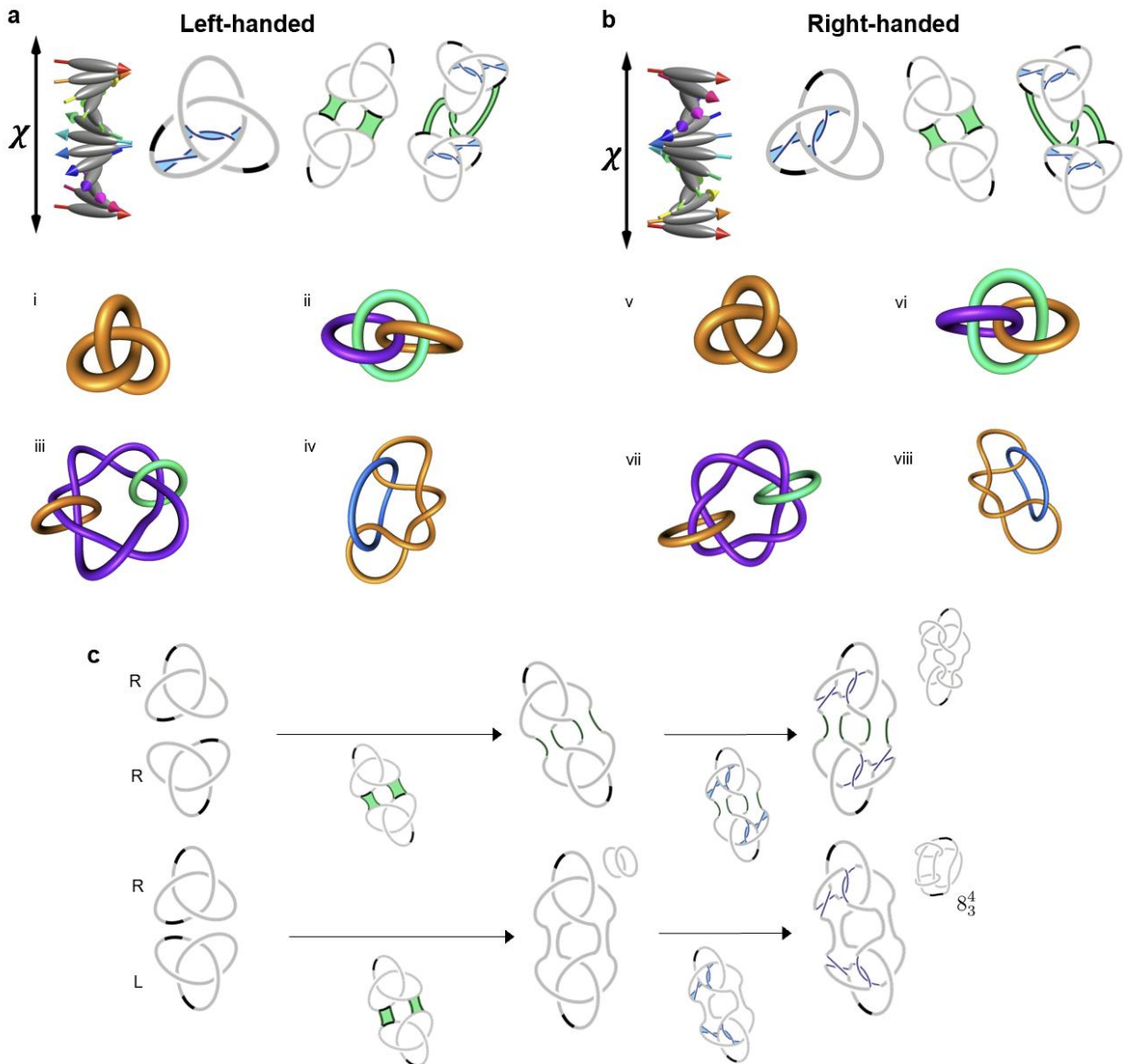

**Extended Data Fig. 10 | Left- and right-handed vortex knots. a-b**, Left and right-handed knot diagrams and their KnotPlot representations summarizing results for a left-handed and a right-handed chiral host LC medium, respectively. Knot diagrams and their reconnection sites are presented to illustrate the one-to-one correspondence between a given knot and the confirmed existence of its mirrored counterpart in a medium of opposite handedness. The direction and handedness of the helical nematic background is shown schematically in the first row. i-iv and v-viii correspond to the simulated left- and right-handed knots, respectively, obtained through re-linking of vortex lines as described above. **c**, Top row describes a typically observed fusion of two right-handed trefoil knots, while the bottom row depicts a hypothetical reconnection between a right- and left-handed trefoil knots that could produce an achiral knot, albeit such opposite-chirality knots so far could not be stabilized next to one another in left- or right-handed or achiral nematic LCs. Letters "L" and "R" denote left- or right-handedness of respective knots.

**Supplementary Video Captions**

**Supplementary Video 1 | Fusion of two heliknotons.** Video shows a pair of heliknotons undergoing reconnections visualized by both director-orientation-based colored vortices and ribbons of splay-bend deformations in the helical axis field around the vortices. The simplified schematics of topology transformations are shown in the top row of the inset in the center. The bottom row of the inset depicts the color scheme corresponding to $\mathbb{S}^2$ and visualizing director orientations, as well as the scheme defining the ribbons of splay-bend deformations in the helical axis field.

**Supplementary Video 2 | Heliknoton fusion starting from a linear array.** A POM video of an array of initially 6 separate heliknotons hybridizing into a fused $Q = 4$ tetramer (left) and $Q = 2$ dimer (right). The scale bar represents 10 µm; crossed double arrows show the orientations of the crossed polarizers.

**Supplementary Video 3 | Voltage-induced reconnection of two heliknotons.** Video shows the reconnections of two heliknotons. The heliknotons initially have the separation vector orthogonal to the helical axis, however, increasing the voltage displaces the heliknotons vertically causing them to fuse together. This fusion and further evolution of the multi-component link topology through reconnections is shown schematically in the central inset's top row, whereas the bottom row illustrates the color schemes depicting director orientations and the local vortex winding numbers. The final 4-component link obtained after transformation from a 3-component link (with the intermediate state in the form of a graph seen in this video), is shown Fig. 2g.

**Supplementary Video 4 | Fission and fusion of knots approaching along $\chi_0$ and z-axis.** Video shows transformations between a pair of heliknotons separated along the far-field helical

axis $\chi_0$ (along z-axis) switching between two separated trefoil knots and the connected sum of the two trefoils. The simplified diagrams of the states before and after reconnecting are shown in the center inset's top row; the bottom row describes the schemes for visualizing colored director orientations (left) and splay-bend regions in the helical axis field around the vortices.

**Supplementary Video 5 | Repeated fusion and fission of two heliknotons.** Video shows a pair of heliknotons having separation vector tilted relative to the far-field helical axis (and z-axis). The simplified schematics of the knots obtained and respective color schemes (similar to ones used in previous videos) are shown in the central inset.

**Supplementary Video 6 | Fusion of two heliknoton trimers.** The video shows two heliknoton trimers, separated along the far-field helical axis, undergoing complex reconnections and forming metastable graphs visualized by both director-orientation-based colored vortices and ribbons of splay-bend around the vortices. The simplified schematics of the knots and respective color schemes (similar to ones used in previous videos) are shown in the inset in the center.

**Supplementary Video 7 | Reconnections within a heliknoton lattice upon switching voltage.** A numerically simulated POM video of a lattice of heliknotons hybridizing into a $Q = 18$ graph. The scale bar represents 10 µm; the crossed double arrows show orientations of crossed polarizers.

**Supplementary Video 8 | Experimental videos depicting fusion of within heliknoton lattices.** The video shows two separate heliknoton lattices with net total Hopf indices $Q = 33$ (left) and $Q = 16$ (right) perturbed from equilibrium by increasing the voltage and subsequently transforming into complex graphs vortices. The scale bar represents 10 µm; the crossed double arrows show the orientations of the crossed polarizers. The experimental details are provided in the captions of Extended Data Fig. 3a,c.

**Supplementary Video 9 | Manipulation of reconnection sites using laser tweezers.** Video depicts two scenarios where heliknoton dimers are manipulated with laser tweezers to hybridize with a fused linear array of heliknotons that were hybridized prior to the beginning of the video. The scale bar represents 10 μm and the double arrows show the orientations of the crossed polarizers. The experimental details are provided in the caption of Extended Data Fig. 4.

**Supplementary Video 10 | Relinkings that involve trimers and tetramers of heliknotons.** The video shows three heliknotons (left) and four heliknotons (right) inter-transforming between different states, including graphs and multicomponent links of dichiralation vortices. Numbers in the bottom right of the knot diagrams represent the Hopf indices for each knot or link visualized by both director-orientation-based coloring of knotted vortices and ribbons depicting splay-bend deformations of the helical axis field around the vortices (see also Extended Data Fig. 6). The simplified schematics of the knots and respective color schemes (similar to ones used in previous videos) are shown in the inset in the center.